\documentclass[aps,preprint,showpacs,showkeys]{revtex4}
\usepackage{amsmath}
\usepackage{bm}

\input{tcilatex}
\begin{document}

\title{An alternative solution of Diatomic Molecules}
\author{\"{O}zg\"{u}r \"{O}ztemel, Eser Ol\u{g}ar}
\affiliation{University of Gaziantep, Engineering of Physics Department, Gaziantep/TURKEY}
\email{ozguroztemel@hotmail.com, olgar@gantep.edu.tr}
\date{\today }

\begin{abstract}
The spectrum of $r^{-1}$ and $r^{-2}$ type potentials of diatomic molecules
in radial Schr\"{o}dinger equation are calculated by using the formalism of
asymptotic iteration method. The alternative method is used to solve
eigenvalues and eigenfunctions of Mie potential, Kratzer-Fues potential,
Coulomb potential, and Pseudoharmonic potential by determining the $\alpha ,$
$\beta ,$ $\gamma $ and $\sigma $ parameters.
\end{abstract}

\maketitle

\section{Introduction}

In the last few decades, there has been raised a great deal of interest in
many branches of physics in order to calculate the energy eigenvalues and
eigenfunctions of diatomic molecules \cite{1}. The most important analytical
methods which are have been used in literature to solve these studies are
supersymmetry (SUSY) \cite{2}, Nikiforov-Uvarov (NU) method \cite{3},
Pekeris approximation \cite{4}, variational method \cite{5}, hypervirial
perturbation method \cite{6}, shifted 1/N expansion (SE) and the modified
shifted 1 /N expansion (MSE) \cite{7}, exact quantization rule (EQR) \cite{8}%
, perturbative formalism \cite{9,10}, polynomial solution \cite{11}, wave
function ansatz method \cite{12}, path integral \cite{13}, Lie algebraic
method \cite{14,15}, Fourier Grid Hamiltonian method \cite{16}, and
asymptotic iteration method (AIM) \cite{ciftci1} to solve the radial Schr%
\"{o}dinger equation exactly.

Among these methods, AIM which was developed by H.Ciftci in 2003 to solve
the Schr\"{o}dinger like second order differential equation, has been used
in many field of physics due to simplicity in obtaining the energy
eigenvalues and corresponding eigenfunctions \cite{olgar,boztosun}. To apply
the method, the asymptotic wavefunction form should be proposed after
substituting the potential in Schr\"{o}dinger equations. Then the AIM is
applied to calculate the spectrum of potentials. But in this study, we
propose an asymptotic wavefunction to the radial Schr\"{o}dinger equation
before substituting the potential function. This yields to write a general
asymptotic form of Schr\"{o}dinger equation which is amenable to apply the
termination condition in AIM. The transformed radial Schr\"{o}dinger
equation is applied to Mie potential, Kratzer-Fues potential, Coulomb
potential, and Pseudoharmonic potential. The energy eigenvalues and
eigenfunctions satisfy the corresponding results in literature.

The organization of this study is as follows: the general Hamiltonian for
diatomic is introduced in the Section II. A\ general description of AIM is
outlined in Section III. Section IV is devoted to proposed form of AIM with
applications. Finally, Section V is devoted to a conclusion.

\section{Formalism of the Proposed Method for Diatomic Molecules}

The Hamiltonian operator for diatomic molecules with spherically symmetric
potential which means depends only on $r$; separation distance between two
molecules, not $\theta $ or $\phi $, takes the form \cite{Griffiths}%
\begin{equation}
\hat{H}=\left\{ \frac{1}{r}\frac{d}{dr}\left( r^{2}\frac{d}{dr}\right) +%
\frac{\widehat{L}^{2}}{2\mu r^{2}}+V\left( r\right) \right\}   \label{e1}
\end{equation}%
\bigskip where $V$ is the potential, $\widehat{L}$ is the angular momentum
operator and $\mu =m_{1}m_{2}/m_{1}+m_{2}$ is called the reduced mass; $m_{1}
$ and $m_{2}$ are the mass of molecules 1 and 2, respectively. By using the
separation of variable for wavefunction%
\begin{equation}
\Psi \left( r,\theta ,\phi \right) =R_{n\ell }\left( r\right) Y_{\ell
}^{m}\left( \theta ,\phi \right)   \label{e2}
\end{equation}

The Schr\"{o}dinger equation by using Eq.(\ref{e1}) and Eq.(\ref{e2}) turns
into%
\begin{equation}
\left\{ \frac{1}{r}\frac{d}{dr}\left( r^{2}\frac{d}{dr}\right) +\frac{\ell
\left( \ell +1\right) \hbar ^{2}}{2\mu r^{2}}+V\left( r\right) \right\}
R_{n\ell }\left( r\right) =E_{n}R_{n\ell }\left( r\right)   \label{e3}
\end{equation}%
where $E$ is the energy, $n$ is the radial quantum number, $\ell $ is the
angular momentum quantum number and $Y_{\ell }^{m}\left( \theta ,\phi
\right) $ is an eigenfunction of $\widehat{L}^{2}$ with eigenvalue $\ell
\left( \ell +1\right) \hbar ^{2}$. Moreover, the angular part of the
wavefunction $Y_{\ell }^{m}$ can be canceled from this equation because all
terms contained one such factor, $r$. The angular momentum $L$ is conserved
that will not change with time for such potentials. This reduces the
Hamiltonian to one dimensional effective system for the radial part of the
wavefunction $R_{n\ell }$ as seen in Eq.(\ref{e3})\cite{Zettilli}.

After rearranging the Eq.(\ref{e3}), we get%
\begin{equation}
\frac{d^{2}R_{n\ell }\left( r\right) }{dr^{2}}+\frac{2}{r}\frac{dR_{n\ell
}\left( r\right) }{dr}+\frac{2\mu }{\hbar ^{2}}\left[ E_{n}-V\left( r\right)
-\frac{\ell \left( \ell +1\right) \hbar ^{2}}{2\mu r^{2}}\right] R_{n\ell
}\left( r\right) =0  \label{e4}
\end{equation}

If we make the following substitution%
\begin{equation}
\rho =2\alpha r^{k}  \label{e5}
\end{equation}%
where $\alpha $ and $k$ are constants. In terms of this variable the Eq.(\ref%
{e4}) becomes,

\begin{equation*}
\frac{d^{2}R_{n\ell }\left( \rho \right) }{d\rho ^{2}}+\frac{1}{\rho }\left( 
\frac{k+1}{k}\right) \frac{dR_{n\ell }\left( \rho \right) }{d\rho }
\end{equation*}

\begin{equation}
+\frac{2\mu }{\hbar ^{2}}\left( \frac{1}{4\alpha ^{2}k^{2}}\right) \left( 
\frac{\rho }{2\alpha }\right) ^{-\frac{2(k-1)}{k}}\left[ E_{n}-V\left(
r\right) -\frac{\ell \left( \ell +1\right) \hbar ^{2}}{2\mu r^{2}}\right]
R_{n\ell }\left( \rho \right) =0  \label{e6}
\end{equation}%
whose solution gives the energy levels of the system by using the Laguerre
functions. The wavefunction $R_{n\ell }\left( \rho \right) $ can be given by
in this form \cite{Gasiorowicz}

\begin{equation}
R\left( \rho \right) =\rho ^{\gamma }\exp [-\frac{\rho }{2}]G(\rho )
\label{e7}
\end{equation}%
where $\gamma $ is a constant and $G(\rho )$ represents Laguerre function.
We substitute this equation into the Eq.(\ref{e6}) and make some algebra to
get the equation for $G(\rho )$. The general expression for this function
can be written as

\begin{equation*}
G^{\prime \prime }(\rho )+\left[ \left( 2\gamma +\frac{k+1}{k}\right) \frac{1%
}{\rho }-1\right] G^{\prime }(\rho )
\end{equation*}

\begin{equation*}
+\left\{ \frac{1}{4}-\left( \gamma +\frac{k+1}{2k}\right) \frac{1}{\rho }%
+\gamma (\gamma -1+\frac{k+1}{k})\frac{1}{\rho ^{2}}\right\} G\left( \rho
\right)
\end{equation*}

\begin{equation}
+\frac{2\mu }{\hbar ^{2}}\left( \frac{1}{4\alpha ^{2}k^{2}}\right) \left( 
\frac{\rho }{2\alpha }\right) ^{-\frac{2(k-1)}{k}}\left[ E_{n}-V\left(
r\right) -\frac{\ell \left( \ell +1\right) \hbar ^{2}}{2\mu r^{2}}\right]
G\left( \rho \right) =0  \label{e8}
\end{equation}

If $k=1$ or $\rho =2\alpha r$, the general expression, Eq.(\ref{e8}) becomes

\begin{equation*}
G^{\prime \prime }(\rho )+\left[ \frac{2\left( \gamma +1\right) }{\rho }-1%
\right] G^{\prime }(\rho )+
\end{equation*}

\begin{equation}
\left\{ \frac{1}{4}-\frac{\left( \gamma +1\right) }{\rho }+\frac{\gamma
(\gamma +1)}{\rho ^{2}}+\frac{2\mu }{\hbar ^{2}}\left( \frac{1}{4\alpha ^{2}}%
\right) \left[ E_{n}-V\left( r\right) -\frac{\ell \left( \ell +1\right)
\hbar ^{2}}{2\mu r^{2}}\right] \right\} G\left( \rho \right) =0  \label{e9}
\end{equation}

If $k=2$ or $\rho =2\alpha r^{2}$, the general expression, Eq.(\ref{e8})
becomes

\begin{equation*}
G^{\prime \prime }(\rho )+\left[ \left( 2\gamma +\frac{3}{2}\right) \frac{1}{%
\rho }-1\right] G^{\prime }(\rho )+
\end{equation*}

\begin{equation}
\left\{ \frac{1}{4}-\frac{\left( \gamma +\frac{3}{4}\right) }{\rho }+\frac{%
\gamma (\gamma +\frac{1}{2})}{\rho ^{2}}+\frac{\mu }{\hbar ^{2}}\left( \frac{%
1}{4\alpha \rho }\right) \left[ E_{n}-V\left( r\right) -\frac{\ell \left(
\ell +1\right) \hbar ^{2}}{2\mu r^{2}}\right] \right\} G\left( \rho \right)
=0  \label{e10}
\end{equation}

Now we will find the analytical solutions of the Eq.(\ref{e8}) for diatomic
potentials with any angular momentum. The Eq.(\ref{e8}) can be transformed
into a second order differential equation form for the corresponding
potentials. The transfomed differential equations are solved within the
framework of the asymptotic iteration method. We apply our formalism to
several important diatomic potentials and obtain the energy eigenvalues and
the corresponding eigenfunctions \cite{h.akcay}.

\section{Asymptotic Iteration Method}

The starting point of AIM is to consider the following second order
homogeneous differential equation:%
\begin{equation}
y^{\prime \prime }(x)=\lambda _{0}y^{\prime }(x)+s_{0}y(x),  \label{a1}
\end{equation}%
where $\lambda _{0}$ and $s_{0}$ are functions and $y^{\prime }(x)$ and $%
y^{\prime \prime }(x)$ denotes derivative of $y$ with respect to $x$. It is
easy to show that $(n+2)^{th}$ derivative of the function $y(x)$ can be
written as%
\begin{equation}
y^{(n+2)}(x)=\lambda _{n}y^{\prime }(x)+s_{n}y(x)  \label{a2}
\end{equation}%
where $\lambda _{n}$ and $s_{n}$ are given by the recurrence relations%
\begin{equation}
\lambda _{n}=\lambda _{n-1}^{\prime }+s_{n-1}+\lambda _{n-1}\lambda _{0}%
\text{,\quad\ }s_{n}=s_{n-1}^{\prime }+\lambda _{n-1}s_{0}  \label{a3}
\end{equation}

If we have, for sufficiently large $n$, 
\begin{equation}
\frac{\lambda _{n}}{s_{n}}=\frac{\lambda _{n-1}}{s_{n-1}}=\alpha (x)
\label{a4}
\end{equation}%
then the solution of Eq. (\ref{a1}) can be written as below%
\begin{equation}
y(x)=\exp \left( -\int^{x}\alpha dt\right) \left[ C_{1}+C_{2}\int^{x}\exp
\left( \int^{t}\left( \lambda _{0}+2\alpha \right) d\tau \right) dt\right]
\label{a5}
\end{equation}

In calculating the parameters in Eq. (\ref{a3}), for $n=0$, we take the
initial conditions as $\lambda _{-1}=1$ and $s_{-1}=0$ \cite{fernandez} and $%
\Delta _{n}(x)=0$ for%
\begin{equation}
\Delta _{n}(x)=\lambda _{n}(x)s_{n-1}(x)-\lambda _{n-1}(x)s_{n}(x)
\label{a6}
\end{equation}%
where $\Delta _{n}(x)$ is the termination condition method in Eq. (\ref{a4}).

In order to find the corresponding energy eigenfunctions, the following wave
function generator is used (\cite{m.aygun},\cite{o.bayrak})

\begin{equation}
f_{n}(x^{\prime })=\dint \frac{\lambda _{n}(x^{\prime })}{s_{n}(x^{\prime })}%
dx^{\prime }  \label{a7}
\end{equation}%
where $x^{\prime }$ is a variable and n is the radial quantum number.

\section{Application of Proposed Method in Diatomic Potentials}

\subsection{Mie Potential}

Mie potential is the intermolecular pair potential which was proposed by
Gustav Mie (1903) who was first to introduce an attractive term and a
repulsive one. The attractive term is the van der Waals interaction
potential which varies with the inverse power of the distance between
molecules \cite{dickerson}. The superposition of these terms produces an
effective potential pocket and the form of this pocket is very important for
the correct energy eigenvalues. For the Mie potential $r\rightarrow 0$, $%
V(r)\rightarrow \infty $ because there is internuclear repulsion. As $%
r\rightarrow \infty $, $V(r)\rightarrow 0$,i.e. the molecule decomposes \cite%
{o.bayrak}.

\bigskip Generally one can define the Mie-type potential as (\cite{mie1},%
\cite{mie2})

\begin{equation}
V(r)=\epsilon \left[ \frac{p}{q-p}\left( \frac{a}{r}\right) ^{q}-\frac{q}{q-p%
}\left( \frac{a}{r}\right) ^{p}\right]  \label{e11}
\end{equation}%
where $\epsilon $ is the interaction energy between two atoms in a molecular
system at distance $a$, $q$ and $p$ are constants which $q\succ p$ is always
satisfied. The solution of the one-dimensional Mie potential with $q=2p$
combination by choosing the special case $p=1$ takes the following form

\begin{equation}
V(r)=V_{0}\left[ \frac{1}{2}\left( \frac{a}{r}\right) ^{2}-\left( \frac{a}{r}%
\right) \right]  \label{e12}
\end{equation}%
where $V_{0}=2\epsilon p$ is the dissociation energy and a is the
equilibrium internuclear distance. Inserting this potential equation into
the general expression for $k=1$, the Eq.(\ref{e9}) can be written%
\begin{equation*}
G^{\prime \prime }(\rho )+\left[ \frac{2\left( \gamma +1\right) }{\rho }-1%
\right] G^{\prime }(\rho )+\left\{ \frac{1}{4}-\frac{\left( \gamma +1\right) 
}{\rho }+\frac{\gamma (\gamma +1)}{\rho ^{2}}\right\} G\left( \rho \right)
\end{equation*}

\begin{equation}
+\left\{ \frac{2\mu E_{n}}{\hbar ^{2}4\alpha ^{2}}-\frac{\mu V_{0}}{\hbar
^{2}}\frac{a^{2}}{4\alpha ^{2}r^{2}}+\frac{2\mu V_{0}}{\hbar ^{2}}\frac{a}{%
4\alpha ^{2}r}-\frac{\ell \left( \ell +1\right) }{4\alpha ^{2}r^{2}}\right\}
G\left( \rho \right) =0  \label{e13}
\end{equation}

Rearranging the Eq.(\ref{e13}) with constants, we get%
\begin{equation*}
G^{\prime \prime }(\rho )+\left[ \frac{2\left( \gamma +1\right) }{\rho }-1%
\right] G^{\prime }(\rho )
\end{equation*}

\begin{equation}
+\left\{ \frac{1}{4}-\frac{\left( \gamma +1\right) }{\rho }+\frac{\beta }{%
\alpha \rho }+\frac{\gamma (\gamma +1)}{\rho ^{2}}-\frac{\ell \left( \ell
+1\right) }{\rho ^{2}}-\frac{\beta a}{\rho ^{2}}-\frac{1}{4}\right\} G\left(
\rho \right) =0  \label{e14}
\end{equation}%
where

\begin{eqnarray*}
\alpha ^{2} &=&-\frac{2\mu E_{n}}{\hbar ^{2}} \\
\beta &=&\frac{\mu V_{0}a}{\hbar ^{2}}
\end{eqnarray*}

To find the $\gamma $ term, we make the following assumption

\begin{equation}
\frac{\gamma (\gamma +1)}{\rho ^{2}}-\frac{\left[ \ell \left( \ell +1\right)
+\beta a\right] }{\rho ^{2}}=0  \label{e15}
\end{equation}%
and the solutions of this equation are obtained as

\begin{equation}
\gamma _{1,2}=\frac{1}{2}\left( -1\mp \sqrt{1+4\sigma }\right) .  \label{e16}
\end{equation}%
where $\sigma =\ell \left( \ell +1\right) +\beta a$ is a constant. The
positive term must be chosen for appropriate solutions of eigenvalues \cite%
{mie1}. For simplicity we use $\gamma +1=\sigma /\rho $, and we get the
general expression in this form

\begin{equation}
G^{\prime \prime }(\rho )=\left[ 1-\frac{2\sigma }{\gamma \rho }\right]
G^{\prime }(\rho )+\left\{ \frac{\sigma }{\gamma \rho }-\frac{\beta }{\alpha
\rho }\right\} G\left( \rho \right) =0  \label{e17}
\end{equation}

The Eq.(\ref{e17}) is a homogeneous linear second-order differential
equation, it can be solved by the AIM where%
\begin{equation*}
\lambda _{0}=1-\frac{2\sigma }{\gamma \rho }
\end{equation*}
\begin{equation}
s_{0}=\frac{\sigma }{\gamma \rho }-\frac{\beta }{\alpha \rho }  \label{e18}
\end{equation}

If we use the termination condition of the AIM in Eq.(\ref{a6}), the energy
eigenvalues are obtained as follows

\begin{eqnarray*}
\lambda _{1}s_{0}-s_{1}\lambda _{0} &=&0\implies E_{0}=-\frac{\hbar
^{2}\beta ^{2}\gamma ^{2}}{2\mu \sigma ^{2}} \\
\lambda _{2}s_{1}-s_{2}\lambda _{1} &=&0\implies E_{1}=-\frac{\hbar
^{2}\beta ^{2}\gamma ^{2}}{2\mu \left( \gamma +\sigma \right) ^{2}} \\
\lambda _{3}s_{2}-s_{2}\lambda _{3} &=&0\implies E_{2}=-\frac{\hbar
^{2}\beta ^{2}\gamma ^{2}}{2\mu \left( 2\gamma +\sigma \right) ^{2}} \\
&&...............
\end{eqnarray*}%
which can be generalized as following

\begin{equation}
E_{n}=-\frac{\hbar ^{2}}{2\mu }\left( 2\beta \right) ^{2}[2n+1+\sqrt{%
1+4\sigma }]^{-2}  \label{e19}
\end{equation}%
Here, n denotes the radial quantum number; $n=0,1,2,...$. Using the Eq.(\ref%
{a7}), we can write the corresponding eigenfunctions

\begin{eqnarray*}
f_{0}(\rho ) &=&1 \\
f_{1}(\rho ) &=&2+2\gamma -\rho \\
f_{2}(\rho ) &=&6+4\gamma ^{2}+10\gamma -\gamma \rho -6\rho +\rho ^{2} \\
f_{3}(\rho ) &=&24+52\gamma +36\gamma ^{2}+8\gamma ^{3}-36\rho -42\gamma
\rho -12\gamma ^{2}\rho +12\rho ^{2}+6\gamma \rho ^{2}-\rho ^{3} \\
&&...............
\end{eqnarray*}

It is understood from the results given above that we can write the general
formula for $f_{n}(\rho )$ as Laguerre function using Eq.(\ref{e16})

\begin{equation}
f_{n}\left( \rho \right) =L_{n}^{\sqrt{1+4\sigma }}\left( \rho \right)
\label{e20}
\end{equation}

Thus we can write the radial wavefunction as below substituting the results
in Eq.(\ref{e5}) and Eq.(\ref{e20}) into the Eq.(\ref{e7})

\begin{equation}
R_{n\ell }\left( \rho \right) =\rho ^{\frac{1}{2}\left( -1+\sqrt{1+4\sigma }%
\right) }\exp [-i\varepsilon r]L_{n}^{\sqrt{1+4\sigma }}\left( 2i\varepsilon
r\right)  \label{e21}
\end{equation}%
where $\varepsilon ^{2}=2\mu E/\hbar ^{2}$.They are exactly same as with the
eigenvalue and eigenfunction equations obtained in \cite{mie1}

\subsection{Kratzer-Fues potential}

The Kratzer-Fues potential is a Mie type potential and it is modified by
adding a $D_{e}$ term to potential in the Eq.(\ref{e12}). A new type of this
potential is called the modified Kratzer potential and it is given by (\cite%
{kratzer1},\cite{mie2},\cite{pseu1}) 
\begin{equation}
V(r)=D_{e}\left[ \frac{r-r_{e}}{r}\right] ^{2}=D_{e}\left[ 1-2\frac{r_{e}}{r}%
+\frac{r_{e}^{2}}{r^{2}}\right]  \label{m1}
\end{equation}%
where $D_{e}$ is the interaction energy between two atoms in a molecular
system at equilibrium distance $r_{e}$. And the general expression for $k=1$
in Eq.(\ref{e9}) can be written with this potential

\begin{equation*}
G^{\prime \prime }(\rho )+\left[ \frac{2\left( \gamma +1\right) }{\rho }-1%
\right] G^{\prime }(\rho )
\end{equation*}

\begin{equation*}
+\left\{ \frac{1}{4}-\frac{\left( \gamma +1\right) }{\rho }+\frac{\gamma
(\gamma +1)}{\rho ^{2}}\right\} G\left( \rho \right)
\end{equation*}
\begin{equation}
+\left\{ \frac{2\mu \left( E_{n}-D_{e}\right) }{\hbar ^{2}4\alpha ^{2}}-%
\frac{\mu D_{e}}{\hbar ^{2}}\frac{r_{e}^{2}}{4\alpha ^{2}r^{2}}+\frac{2\mu
D_{e}}{\hbar ^{2}}\frac{r_{e}}{4\alpha ^{2}r}-\frac{\ell \left( \ell
+1\right) }{4\alpha ^{2}r^{2}}\right\} G\left( \rho \right) =0  \label{m2}
\end{equation}

\bigskip The following equation is obtained by using the Eqs.(\ref{e14},\ref%
{e15},\ref{e16})%
\begin{equation}
G^{\prime \prime }(\rho )=\left[ 1-\frac{2\sigma }{\gamma \rho }\right]
G^{\prime }(\rho )+\left\{ \frac{\sigma }{\gamma \rho }-\frac{\beta }{\alpha
\rho }\right\} G\left( \rho \right) =0  \label{m3}
\end{equation}%
where

\begin{eqnarray*}
\alpha ^{2} &=&-\frac{2\mu \left( E_{n}-D_{e}\right) }{\hbar ^{2}} \\
\beta &=&\frac{2\mu D_{e}r_{e}}{\hbar ^{2}} \\
\gamma &=&\frac{1}{2}\left( -1+\sqrt{1+4\sigma }\right) \\
\sigma &=&\ell \left( \ell +1\right) +\beta r_{e}
\end{eqnarray*}

Now, we reach to a position that the differential equation is suitable for
applying AIM. Therefore, the energy eigenvalues of Eq.(\ref{m3}) should has
a solution in the form of Eq.(\ref{e19})%
\begin{equation}
\varepsilon _{n}=-\left( 2\beta \right) ^{2}[2n+1+\sqrt{1+4\sigma }]^{-2}
\label{m4}
\end{equation}%
where $\varepsilon _{n}^{2}=2\mu \left( E_{n}-D_{e}\right) /\hbar ^{2}$, $%
\varepsilon $ is called energy spectrum. Using the wavefunction generator,
we get the energy eigenfunction as in the Eq.(\ref{e20}). The Eq.(\ref{e5})
and the Eq.(\ref{e7}) give the following formula for the total radial
wavefunction%
\begin{equation}
R_{n\ell }\left( \rho \right) =\rho ^{\frac{1}{2}\left( -1+\sqrt{1+4\sigma }%
\right) }\exp [-\frac{\rho }{2}]L_{n}^{\sqrt{1+4\sigma }}\left( \rho \right)
\label{m5}
\end{equation}%
which is exactly same as with the eigenvalue equation obtained in \cite%
{kratzer1} through a proper choice of parameters.

\subsection{Coulomb potential}

The coulomb potential is an effective pair potential that describes the
electrostatic interaction between electrically charged particles. This
potential is given by%
\begin{equation}
V(r)=-\frac{kZe^{2}}{r}  \label{k1}
\end{equation}%
where $r$ is the distance between two atoms, $e$ is the electron charge, $Z$
is the atomic number and $k=1/4\pi \epsilon _{0}$ is a constant; $\epsilon
_{0}$ is the electrical permittivity of the space. By using the potential,
the same expression for k=1 is obtained as in Eq.((\ref{e17}) 
\begin{equation}
G^{\prime \prime }(\rho )=\left[ 1-\frac{2\sigma }{\gamma \rho }\right]
G^{\prime }(\rho )+\left\{ \frac{\sigma }{\gamma \rho }-\frac{\beta }{\alpha
\rho }\right\} G\left( \rho \right) =0  \label{k2}
\end{equation}%
where

\begin{eqnarray*}
\alpha ^{2} &=&-\frac{2\mu E_{n}}{\hbar ^{2}} \\
\beta &=&\frac{\mu kZe^{2}}{\hbar ^{2}} \\
\gamma &=&\frac{1}{2}\left( -1+\sqrt{1+4\sigma }\right) \\
\sigma &=&\ell \left( \ell +1\right)
\end{eqnarray*}

Here for this potential $\gamma =\ell $ and in general we obtain the energy
eigenvalues by changing the parameters in Eq.(\ref{e19})

\begin{equation}
E_{n}=-\frac{\mu k^{2}Z^{2}e^{4}}{2\hbar ^{2}(n+\ell +1)^{2}}  \label{k3}
\end{equation}%
where $n$ is the radial quantum number and $\ell $ is the angular momentum
quantum number. We obtained the general formula of corresponding
eigenfunctions for the Eq.(\ref{k2}) as Laguerre function in Eq.(\ref{e20}).
Again by changing the parameters we can write the total radial wavefunction%
\begin{equation}
R_{n\ell }\left( \rho \right) =N\rho ^{\ell }\exp [-\frac{\rho }{2}%
]L_{n}^{2\ell +1}\left( \rho \right)  \label{k4}
\end{equation}%
where N is a normalization constant and the total energy eigenvalues in Eq. (%
\ref{k3}) and the total wavefunction in (\ref{k4}) are exactly same as with
the values that obtained in \cite{h.akcay}.

\subsection{Pseudoharmonic potential}

Pseudoharmonic potential can be considered as an intermediate potential
between the harmonic oscillator potential and the Morse-type potentials,
which are more realistic anharmonic potentials. The Pseudoharmonic potential
is one of the exactly solvable potential and defines the real physical
systems that have generally anharmonical properties \cite{pseu2}. It can be
written as \cite{pseu3}

\begin{equation}
V(r)=V_{0}\left[ \frac{r}{r_{0}}-\frac{r_{0}}{r}\right] ^{2}=V_{0}\left( 
\frac{r^{2}}{r_{0}^{2}}+\frac{r_{0}^{2}}{r^{2}}-2\right)   \label{p1}
\end{equation}%
where $V_{0}$ is the dissociation energy and $r_{0}$ is the equilibrium
intermolecular separation. Inserting the potential to the general expression
for $k=2$ in Eq.(\ref{e10}), the equation becomes

\begin{equation*}
G^{\prime \prime }(\rho )+\left[ \left( 2\gamma +\frac{3}{2}\right) \frac{1}{%
\rho }-1\right] G^{\prime }(\rho )
\end{equation*}

\begin{equation*}
+\left\{ \frac{1}{4}-\frac{\left( \gamma +\frac{3}{4}\right) }{\rho }+\frac{%
\gamma (\gamma +\frac{1}{2})}{\rho ^{2}}\right\} G\left( \rho \right)
\end{equation*}

\begin{equation}
+\left\{ \frac{\mu \left( E_{n}+2V_{0}\right) }{\hbar ^{2}4\alpha \rho }-%
\frac{\mu V_{0}}{\hbar ^{2}4\alpha \rho }\frac{r^{2}}{r_{0}^{2}}-\frac{\mu
V_{0}}{\hbar ^{2}4\alpha \rho }\frac{r_{0}^{2}}{r^{2}}-\frac{\ell \left(
\ell +1\right) }{8\alpha \rho r^{2}}\right\} G\left( \rho \right) =0
\label{p2}
\end{equation}

After rearranging this equation with following parameters

\begin{eqnarray*}
\alpha ^{2} &=&\frac{\mu V_{0}}{2\hbar ^{2}r_{0}^{2}} \\
\beta &=&\frac{\mu V_{0}r_{0}^{2}}{2\hbar ^{2}}
\end{eqnarray*}%
we get

\begin{equation*}
G^{\prime \prime }(\rho )+\left[ \left( 2\gamma +\frac{3}{2}\right) \frac{1}{%
\rho }-1\right] G^{\prime }(\rho )
\end{equation*}

\begin{equation}
+\left\{ \frac{1}{4}+\left[ \frac{\mu \left( E_{n}+2V_{0}\right) }{\hbar
^{2}4\alpha }-\gamma -\frac{3}{4}\right] \frac{1}{\rho }-\frac{1}{4}+\left[
\gamma (\gamma +\frac{1}{2})-\beta -\frac{\ell \left( \ell +1\right) }{4}%
\right] \frac{1}{\rho ^{2}}\right\} G\left( \rho \right) =0  \label{p3}
\end{equation}

If we make the following assumption to find the $\gamma $ term

\begin{equation*}
\left[ \gamma (\gamma +\frac{1}{2})-\beta -\frac{\ell \left( \ell +1\right) 
}{4}\right] \frac{1}{\rho ^{2}}=0
\end{equation*}

The solution of this equation gives us the $\gamma =\frac{1}{4}(-1+\sqrt{%
1+16\sigma })$ where $\sigma =\beta +\frac{\ell \left( \ell +1\right) }{4}$.
The general expression becomes for this potential

\begin{equation}
G^{\prime \prime }(\rho )=\left[ 1-\left( 2\gamma +\frac{3}{2}\right) \frac{1%
}{\rho }\right] G^{\prime }(\rho )+\left[ \left( \gamma +\frac{3}{4}-\frac{%
\mu \left( E_{n}+2V_{0}\right) }{2\hbar ^{2}2\alpha }\right) \frac{1}{\rho }%
\right] G\left( \rho \right)  \label{p4}
\end{equation}

By using the termination condition of AIM in Eq.(\ref{a6}) where

\begin{eqnarray*}
\lambda _{0} &=&1-\left( 2\gamma +\frac{3}{2}\right) \frac{1}{\rho } \\
s_{0} &=&\left( \gamma +\frac{3}{4}-\frac{\mu \left( E_{n}+2V_{0}\right) }{%
2\hbar ^{2}2\alpha }\right) \frac{1}{\rho }
\end{eqnarray*}

The eigenvalues of the Eq.(\ref{p4}) is obtained as

\begin{eqnarray*}
\lambda _{1}s_{0}-s_{1}\lambda _{0} &=&0\implies \varepsilon _{0}=\alpha
\left( \frac{3}{2}+2\gamma \right) \\
\lambda _{2}s_{1}-s_{2}\lambda _{1} &=&0\implies \varepsilon _{1}=\alpha
\left( \frac{7}{2}+2\gamma \right) \\
\lambda _{3}s_{2}-s_{2}\lambda _{3} &=&0\implies \varepsilon _{2}=\alpha
\left( \frac{11}{2}+2\gamma \right) \\
&&...............
\end{eqnarray*}%
which can be generalized as

\begin{equation}
\varepsilon _{n}=\alpha \lbrack 2n+1+2(\gamma +\frac{1}{4})]  \label{p5}
\end{equation}%
where $\varepsilon _{n}=\frac{\mu \left( E_{n}+2V_{0}\right) }{2\hbar ^{2}}$
and n is radial quantum number. The corresponding wavefunctions can be found
by using the Eq.(\ref{a7})

\begin{eqnarray*}
f_{0}(\rho ) &=&1 \\
f_{1}(\rho ) &=&3+4\gamma -2\rho \\
f_{2}(\rho ) &=&15+32\gamma +16\gamma ^{2}-16\gamma \rho -20\rho +\rho ^{2}
\\
f_{3}(\rho ) &=&105+284\gamma +240\gamma ^{2}+64\gamma ^{3}-288\gamma \rho
-96\gamma ^{2}\rho -210\rho +84\rho ^{2}+48\gamma \rho ^{2}-8\rho ^{3} \\
&&...............
\end{eqnarray*}%
from these equations we can write the general formula for $f_{n}(\rho )$ as
Laguerre function

\begin{equation*}
f_{n}(\rho )=L_{n}^{\frac{1}{2}\sqrt{1+16\sigma }}
\end{equation*}

Now we can find the total radial wavefunction from Eq.(\ref{e7}) by changing
the parameters

\begin{equation}
R_{n\ell }\left( \rho \right) =\rho ^{\frac{1}{4}(-1+\sqrt{1+16\sigma }%
)}\exp [-\frac{\rho }{2}]L_{n}^{\frac{1}{2}\sqrt{1+16\sigma }}\left( \rho
\right)  \label{p6}
\end{equation}%
which is exactly as same as with the equation obtained in \cite{pseu3}

\section{Conclusion}

In this study, an anlternative method is proposed for the solution of $r^{-1}
$ and $r^{-2}$ type potentials of diatomic molecules in Schr\"{o}dinger
equation. The proposed method not only shows compliance with solutions that
obtained by the asymptotic iteration method (AIM) but also transform to the
second order differential equation form which can be solved by AIM.

By determining the $\alpha ,$ $\beta ,$ $\gamma $ and $\sigma $ parameters
for Mie potential, Kratzer-Fues potential, Coulomb potential, and
Pseudoharmonic potential, the corresponding eigenvalues and eigenfunctions
are calculated exactly. In addition to satisfying the results in
literature,they are obtained very easier than the other solution methods
without using complex algebraic calculation.

The proposed method gives the exact solutions only for the specified type
potentials. By changing the function in Eq .\ref{e5}, the general form
radial equation can be applied to the other potentials such as exponential,
trigonometric, etc.

\section{Acknowledgement}

The research was supported by the Research Fund of Gaziantep University
(BAP) and the Scientific and Technological Research Council of TURKEY (T\"{U}%
B\.{I}TAK).

\end{document}